\documentclass[epj]{svjour}
\usepackage{graphicx}
\begin{document}

\title{The spin-1/2 anisotropic Heisenberg-chain in longitudinal and
 transversal magnetic fields: a DMRG study.}

\titlerunning{The anisotropic Heisenberg-chain}
\authorrunning{Felicien~Capraro and Claudius~Gros}


\author{Felicien~Capraro and Claudius~Gros}
\institute{Universit\"at des Saarlandes, Fakult\"at 7.1,
           Postfach 15 11 50, 66041 Saarbr\"ucken, Germany 
           }

\abstract {Using the density matrix renormalization group technique, 
we evaluate the low-energy spectrum (ground state and 
first excited states) of the anisotropic antiferromagnetic 
spin-one-half chain under magnetic fields. We study both
homogeneous longitudinal and transversal fields as well
as the influence of a transversal staggered field on
opening of a spin-gap. We find that only a staggered
transversal field opens a substantial gap.}

\maketitle

\section{Introduction}
Recently, the properties of low-dimensional quantum spin
systems in longitudinal, transversal and/or staggered magnetic
fields have become of considerable interest. To give an example, the 
two-dimensional compound $\rm SrCu_2(BO_3)_2$ is a realization of the 
Shastry-Sutherland model \cite{shastry81},
close to quantum-criticality, with a spin gap of $\rm 31~K$ which
shows magnetization plateaus in an external field \cite{kageyama98b}.

Most spin-1/2 systems show little anisotropies in the magnetic
exchange. The discovery \cite{schmidt96} that
low-dim\-en\-sio\-nal magnetic excitations in the rare-earth 
compound $\rm Yb_4As_3$ can explain the large linear
specific heat coefficient $\gamma$ in this
low-carrier half-metal \cite{ochiai90,kasuya94}
opens the possibility to study in deeper detail
the properties of a rare-earth quan\-tum\--spin\--chain-system with
its enhanced magnetic an\-iso\-tro\-pies \cite{aoki00,shiba00}.

Inelastic neutron scattering experiments \cite{khogi01} 
on $\rm Yb_4As_3$ found a gap to (all)
magnetic excitations opening in the presence of an external
magnetic field, confirming a prediction \cite{schmidt96}
by Schmidt {\it et al} based on an interpretation 
of previous specific heat data \cite{helfrich98}.
Several proposals have been made in order to
explain this very unusual behavior \cite{uimin00}.
The first model \cite{schmidt96}
is based on inter-chain interactions.
The second model \cite{oshikawa99} is based on
the observation that a staggered Dzyaloshinsky-Moriya
(DM) interaction, which generates an effective
staggered g-tensor, is allowed \cite{aoki00,shiba00}
in the $4f$-compound $\rm Yb_4As_3$. It is know
that a staggered g-tensor leads to a gap in an
external field \cite{oshikawa97,affleck99}.
The third model \cite{uimin00}, based on a mean-field
analyze of the anisotropic spin-chain, proposes that a
gap opens in the presence of a uniform transversal magnetic
field.

Here we will analysis the two latter proposals by
a systematic DMRG-studies of the relevant models.
We find that only the effective staggered g-factor-model
is able to explain the field-dependent opening
of a spin-gap in $\rm Yb_4As_3$.

\subsection{Model and method}
The magnetic properties of $\rm Yb_4As_3$, in the
absence of an external magnetic field,
are well described by an antiferromagnetic Heisenberg spin-$1/2$ chain. 
Switching on the external magnetic field, 
experimental data shows the opening of a gap in the low 
energy excitation spectrum.
However, the standard Heisenberg model in an applied field 
remains gapless from zero magnetic field up to the 
saturation magnetization. The anisotropic
Heisenberg model

\begin{eqnarray}
H = \sum_{i=1}^{L-1}   J_{xy}\left(S_i^xS_{i+1}^x + S_i^yS_{i+1}^y\right) 
                     + J_{z} S_i^zS_{i+1}^z\nonumber\\ 
		     - \sum_{i=1}^L h_x S_i^x 
		     + \sum_{i=1}^L h^{stag}_x (-1)^i\, S_i^x 
		     - \sum_{i=1}^L h_z S_i^z
\label{H}
\end{eqnarray}
with an uniform transversal field $h_x$, a staggered
transversal field $h_x^{stag}$ and a longitudinal
field $h_z$ incorporates all features proposed
\cite{uimin00,oshikawa99} to be relevant for $\rm Yb_4As_3$.
The magnetic fields appearing in (\ref{H}) include the
gyromagnetic g-factors.

The staggered transversal field in (\ref{H}) is induced 
by a staggered Dzyaloshinsky-Moriya interaction given by the term
$\sum_i(-1)^i{\bf D}\cdot \left({\bf S}_i \times {\bf S}_{i+1}\right)$.
Setting $D=|{\bf D}|=J_z\sin(2\theta)$ the DM-term
can be eliminated \cite{oshikawa97,shibata01} 
by a rotation around $\bf D$ by an angle $\theta$ leading
to \ $h_x^{stag} = \sin(\theta) h_z$, which can be interpreted
as an effective staggered g-tensor.

We have simulated the anisotropic
Heisenberg model (\ref{H}) using the 
Density Matrix Renormalization Group (DMRG) 
technique \cite{white}. 
We have, in general,
investigated carefully the dependence of the
results on the number $m$ of states kept
in the DMRG calculations. If not stated otherwise,
we have used open boundary conditions,
the finite-system-size
algorithm and extrapolated the finite-size data
to the thermodynamic limit, setting $J_{xy}\equiv1$
in (\ref{H}).

\subsection{Comparison of DMRG and Bethe-Ansatz results}
\begin{figure}
\smallskip
  \includegraphics[width=8cm]{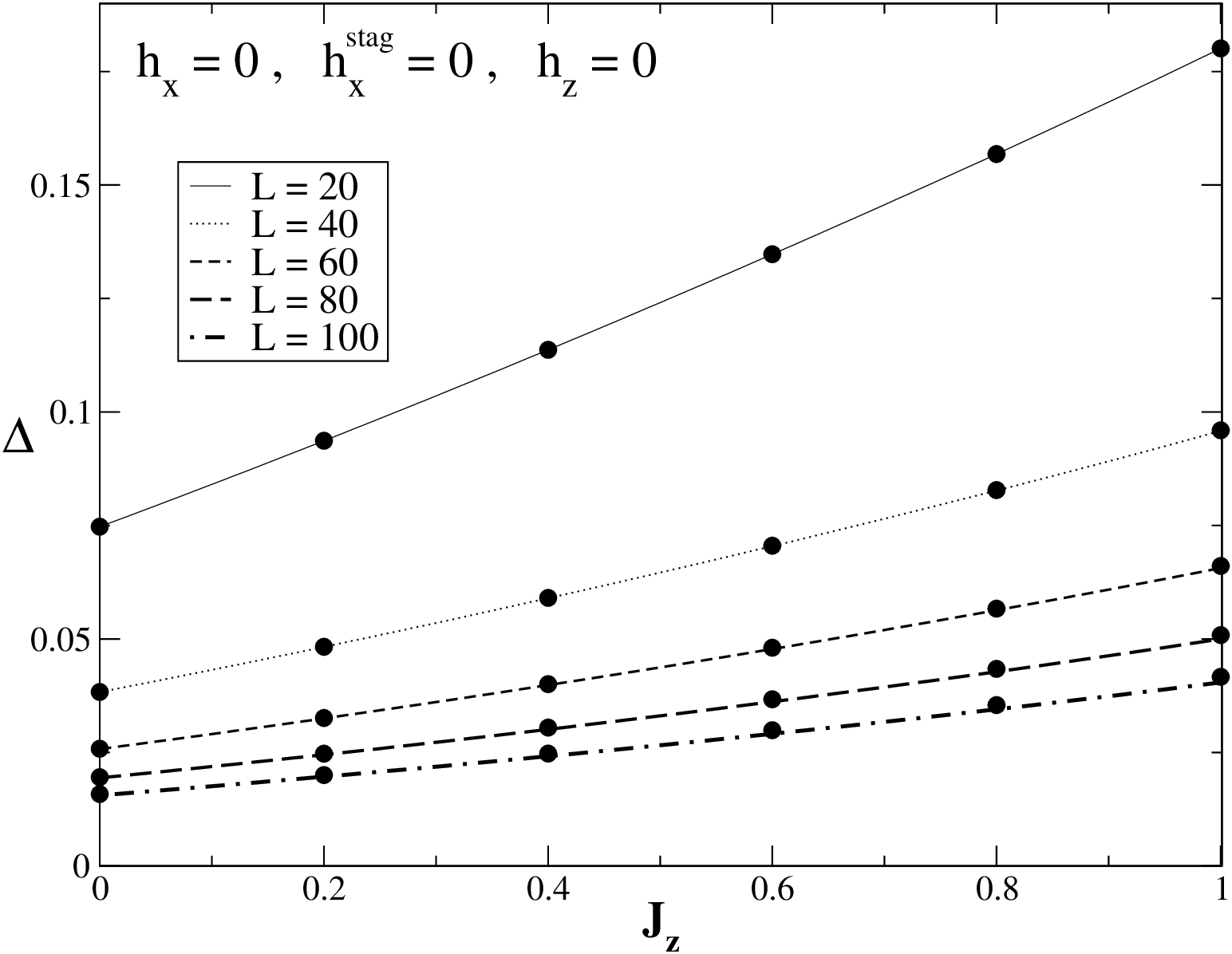}
 \caption{Comparison between DMRG results (filled circles)
 for the finite-size gap $\Delta(L)$, using
 $m = 50$ states, and the Bethe-Ansatz results (lines). 
 The gap is given as a function of the anisotropy $J_z$ in
 the absence of external magnetic fields.
 }
 \label{fig:dmrg-bethe}       
\end{figure}

In order to verify the accuracy of our approach
for the excitation energies
we use the Bethe-Ansatz equations \cite{bethe31,alcaraz87} to evaluate the
gap $\Delta (L)$ for finite chains with size $L$, 
\[
\Delta (L) = E_1(L) - E_0(L)~,
\]
for the anisotropic antiferromagnetic Heisenberg chain 
and then compare it with our DMRG simulations. 
In Fig.~\ref{fig:dmrg-bethe} we plot the finite-size gap 
obtained by using DMRG versus the anisotropy
$(0\leq J_z\leq 1)$, the lines correspond 
to the exact finite-size gap obtained using Bethe-Ansatz. 
Although in this regime the gap in the thermodynamic 
limit is equal to zero, the correspondence between
our DMRG simulations for the gap for finite system 
sizes with the Bethe-Ansatz results is very accurate.


\section{Homogeneous transversal magnetic field}

Uimin {\it et al} proposed, by a mean-field
calculation \cite{uimin00}, that a gap
opens for the spin-1/2 anisotropic Heisenberg chain
\begin{eqnarray}
H = \sum_{i=1}^{L-1} J_{xy}(S_i^xS_{i+1}^x + S_i^yS_{i+1}^y) 
+J_z S_i^zS_{i+1}^z\nonumber\\ 
- h_x\sum_{i=1}^L S_i^x 
\label{model_fulde}
\end{eqnarray}
in the presence of a homogeneous longitudinal field $h_x$.
This mean-field result would then imply that
no further magnetic anisotropies would be needed to
explain the experimentally observed spin-gap
of $\rm Yb_4As_3$.

In spin-wave theory Uimin {\it et al} 
found \cite{uimin00} for the model (\ref{model_fulde})
a gap $\Delta_{SWT}$ consisting of two branches:
\[
\Delta_{SWT} = {\rm min}(\Delta_1,\Delta_2)~,
\]
with $\Delta_1=h_x$ and
\begin{equation}
 \Delta_2 = 
 \sqrt{\frac{2}{3}\left(\frac{1-J_z}{2+Jz}\right)
 \left((2+J_z)^2-\left(\frac{3h_x}{2}\right)^2\right)} ~.
\end{equation}

\begin{figure}
\smallskip
  \begin{center}
  \includegraphics[width=7cm]{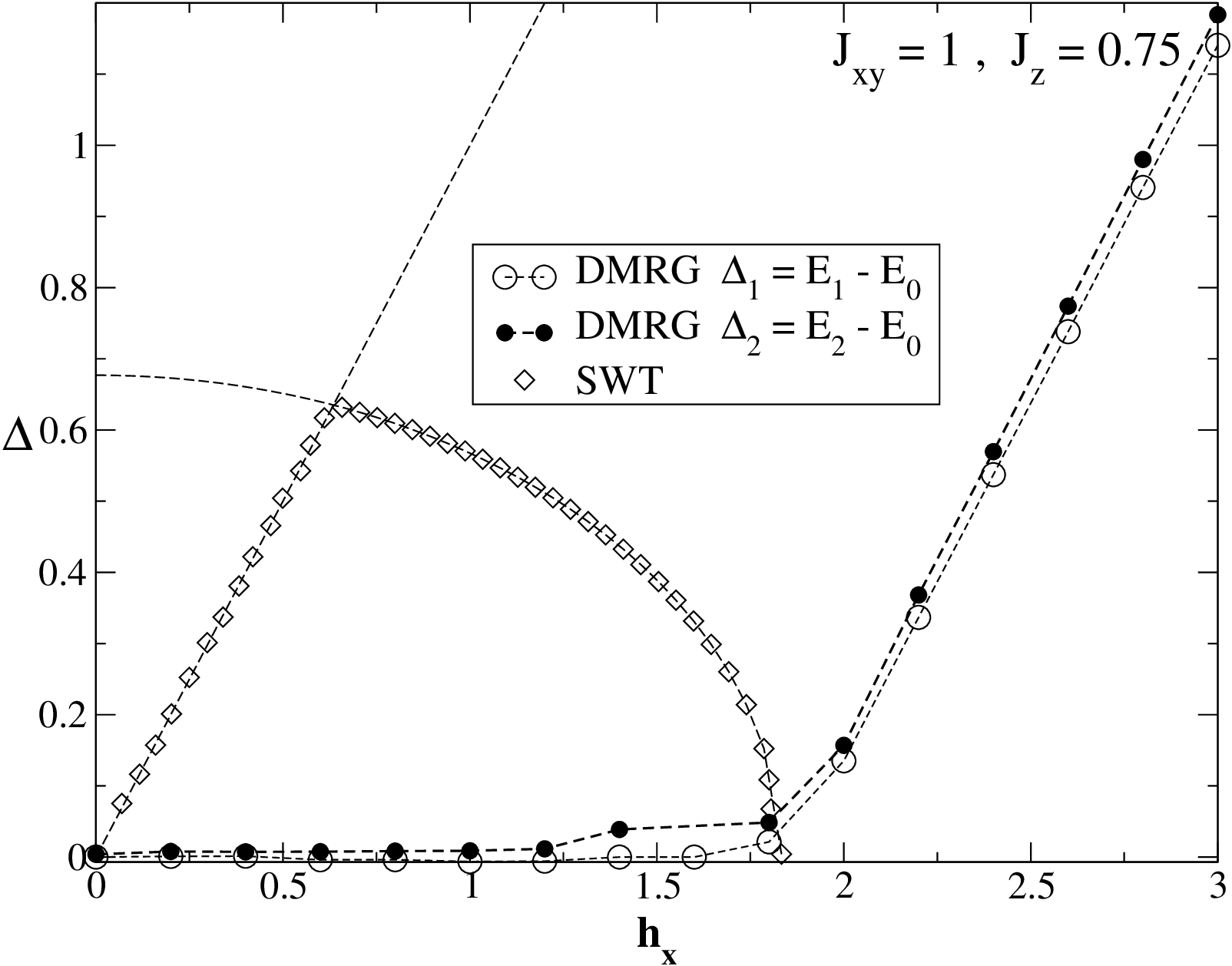}
  \end{center}
 \vspace{-0.5cm}
 \caption{The gap observed for the Heisenberg chain in a 
 transversal magnetic field $h_x$ with $J_{xy} = 1,\, J_z = 0.75$.
The circle symbols come from DMRG simulations $(m = 50)$ 
extrapolated to the thermodynamic limit. 
The diamond symbols represent induced gap calculated 
from spin-wave theory.}
 \label{fig:swt_0.1}       
\end{figure}

\begin{figure}
\smallskip
  \begin{center}
  \includegraphics[width=7cm]{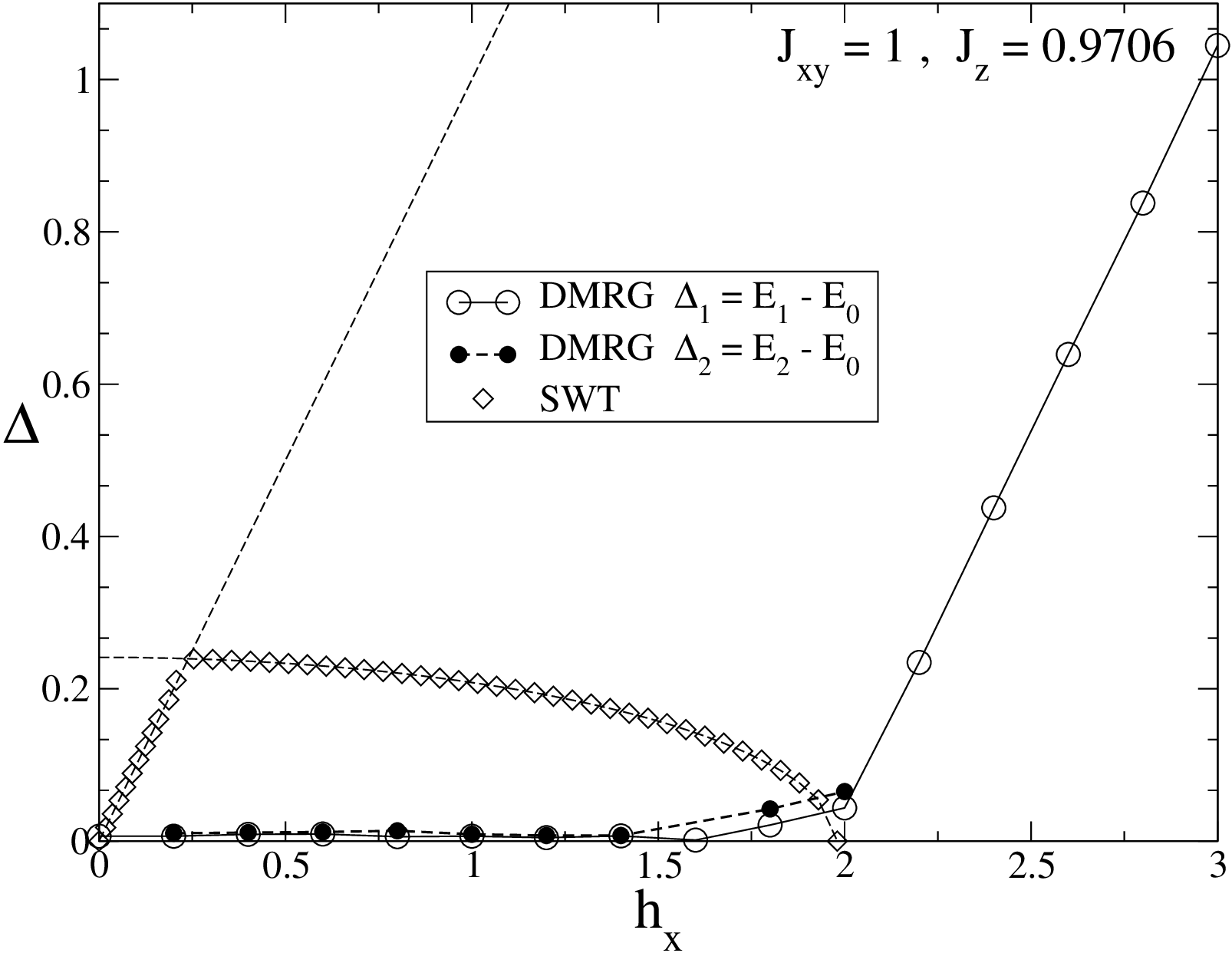}
  \end{center}
 \vspace{-0.5cm}
 \caption{The gap observed for the Heisenberg chain in a 
 transversal magnetic field $h_x$ with $J_{xy} = 1,\, J_z = 0.9706$.
The circle symbols come from DMRG simulations $(m = 50)$ 
extrapolated to the thermodynamic limit. 
The diamonds represent the induced gap calculated 
from spin-wave theory.}
 \label{fig:swt_0.01}       
\end{figure}

Up to a certain magnetic field the value of the gap is almost linear 
and above this value the gap starts to close itself 
following a quadratic form, compare
Fig.~\ref{fig:swt_0.1} and Fig.~\ref{fig:swt_0.01}.   

The physical reasoning for the gap present in the mean-field
results reviewed above is the following: neglecting
the $S_i^x S_{i+1}^x$ coupling term in
the Hamiltonian (\ref{model_fulde}) it becomes identical
to the Ising model in a transversal field, which has
a two-fold degenerate ground-state and a gap.
It has been argued \cite{mori95,dmi02}, that this
reasoning remains valid also for (\ref{model_fulde}).

\begin{figure}
\smallskip
  \begin{center}
  \includegraphics[width=7cm]{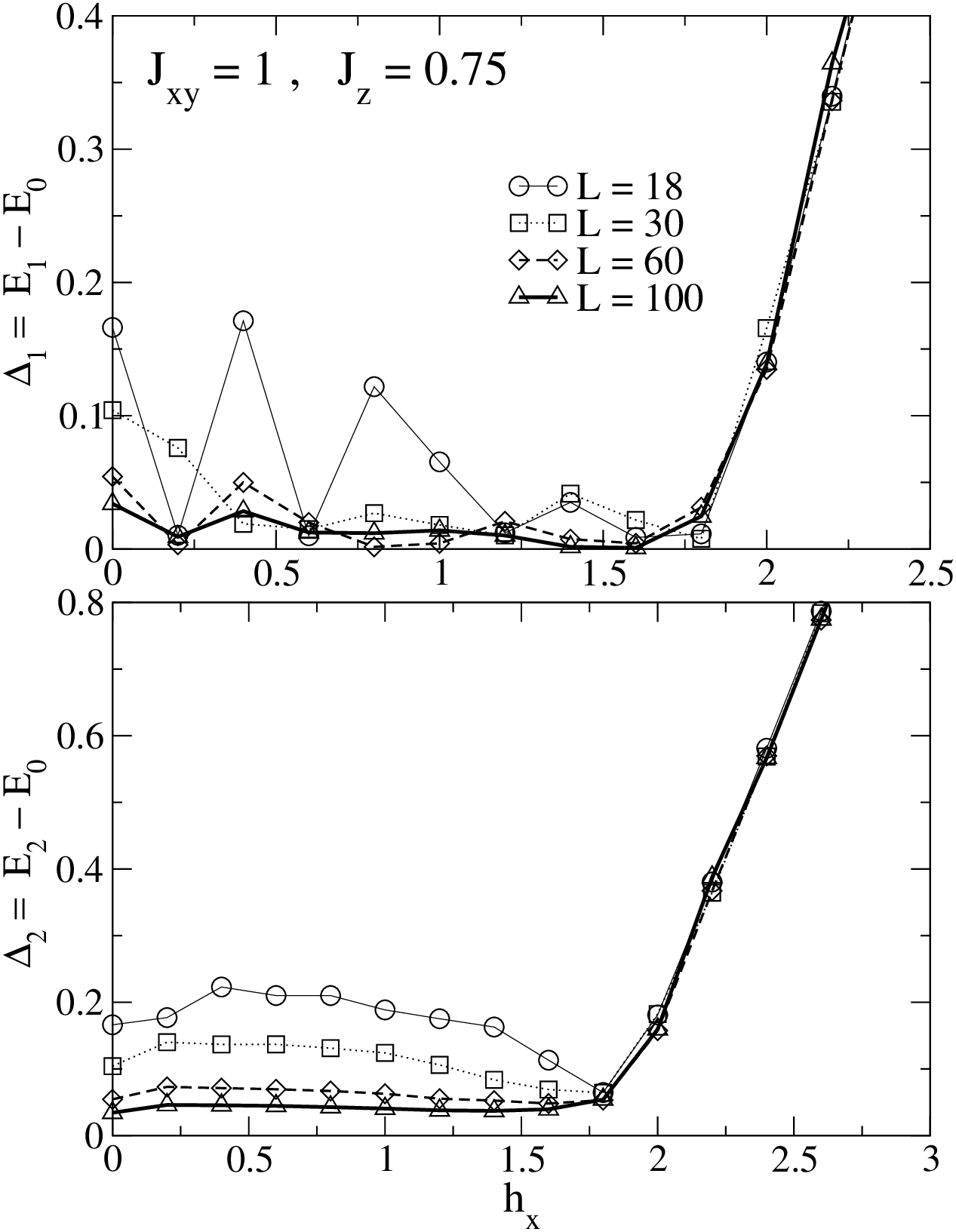}
  \end{center}
 \vspace{-0.5cm}
 \caption{Finite-size gaps observed for the Heisenberg chain 
 of size $L$ in a 
 transversal magnetic field $h_x$ with $J_{xy} = 1,\, J_z = 0.75$.}
 \label{fig:swt_0.1.1}       
\end{figure}

\begin{figure}
\smallskip
  \begin{center}
  \includegraphics[width=7cm]{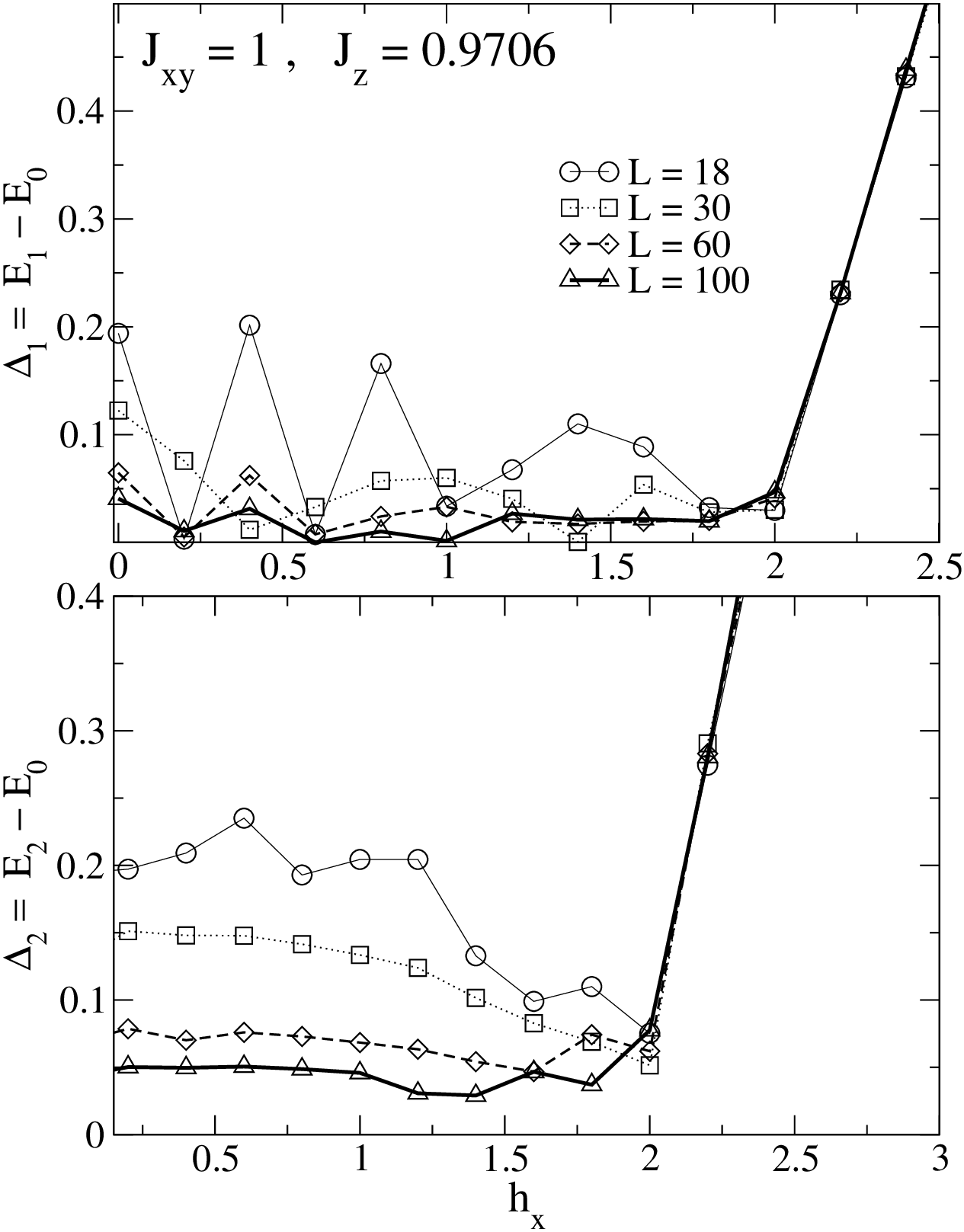}
  \end{center}
 \vspace{-0.5cm}
 \caption{Finite-size gaps observed for the Heisenberg 
 chain of size $L$ in a 
 transversal magnetic field $h_x$ with $J_{xy} = 1,\, J_z = 0.9706$.}
 \label{fig:swt_0.1.2}       
\end{figure}

We compute the energy gap $\Delta(L)$ using the DMRG.
We calculate the ground state and the lowest excited states energies 
$E_i\, (i=0,\,1,\,2\,)$ and form the following energy difference
$\Delta_1~=E_1-E_0$ and $\Delta_2~=E_2-E_0$ respectively 
between the first and the
second excited state with the ground state energy.
Fig.~\ref{fig:swt_0.1.1} and Fig.~\ref{fig:swt_0.1.2} 
show the behavior of $\Delta_i(L)$ 
with the transverse magnetic field $h_x$ for $J_z=0.75,~0.9706$.

\begin{figure}
\smallskip
  \begin{center}
  \includegraphics[width=7cm]{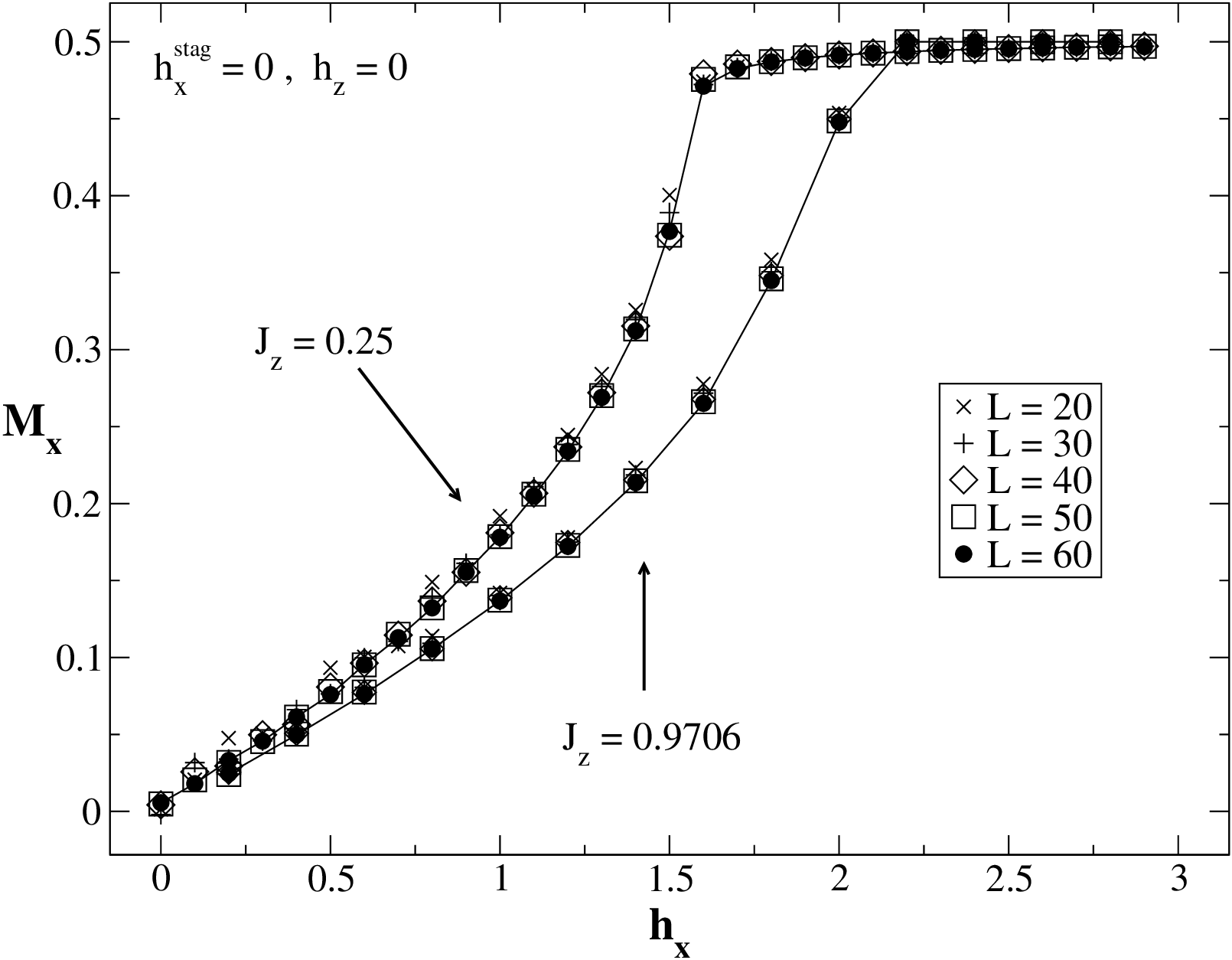}
  \end{center}
 \vspace{-0.5cm}
 \caption{The magnetization $M_x$ along $x$ for a anisotropy
          $J_z = 0.25,\ 0.9706$.
	  Note the saturation of $M_x$
	  above a certain critical field.
	  }
 \label{fig:Mx_L}       
\end{figure}

In the thermodynamic limit the first energy difference $\Delta_1\to 0$ 
in the low magnetic field regime ($h_x\ll2$).
This behavior is in agreement with the fact 
that the GS is doubly degenerated
in this regime when $L\to\infty$. 
Thus to estimate the eventually system-gap in this regime we calculate 
$\Delta_2$. 
$\Delta_2$ shows a more monotonous behavior.
For fixed $h_x$, $\Delta_2(L)$ is decreasing with the system-size $L$
($\Delta_2(100)\leq0.04$).
Fig.~\ref{fig:swt_0.1} and Fig.~\ref{fig:swt_0.01} 
show the extra\-polate gap
for the thermodynamic limit with $J_{xy}\, =\, 1$ and 
$J_z\, =\, 0.75,\, 0.9706$.
We have evaluated also the induced uniform magnetization
$M_x$, which we present in Fig.~\ref{fig:Mx_L}
for $J_z=0.25$ and $J_z=0.9706$ (value which corresponds to 
the small anisotropies in $xy$-plane appropriate for 
$\rm Yb_4 As_3$ \cite{schmidt96}).


We observe:
\begin{description}
\item[i) ]  The DMRG data presented in Fig.~\ref{fig:swt_0.1} and
            Fig.~\ref{fig:swt_0.01} (open circles) show a phase diagram 
	    divided in two regions well separated by a critical 
	    magnetic field, which depends on $J_z$ and which is
            around $h_x\,\simeq\,2$ for $J_z\simeq1$ in agreement with 
		the isotropic case \cite{griffiths}.\\ 
            For $h_x$ below the critical magnetic field the system 
	    appears gapless.  For $h_x$ above the critical magnetic field, 
	    a linear gap opens corresponding to the classical 
            ferromagnetic phase, polarized along $x$-direction.
\item[ii) ] As it appears in the Fig.~\ref{fig:swt_0.1} and
            Fig.~\ref{fig:swt_0.01}, 
	    where we plot the spin-wave theory gap 
            (open diamonds), there is a substantial 
	    disagreement between our 
	    DMRG-sim\-ula\-tion and SWT prediction \cite{uimin00}.
\item[iii) ] We find that the induced magnetization $M_x$ 
             saturate for $h_x$ larger than the critical
	     field, as illustrated in Fig.~\ref{fig:Mx_L} for 
	     $J_z=0.25$ and $J_z=0.9706$
	     (and the staggered magnetization
	     reduces to zero \cite{hieida01}).
	     This explains the opening of a gap
	     linear in $h_x$ in this phase.
\end{description}

Dmitriev {\it et al} have used
scaling arguments \cite{dmi02} and
found for the Hamiltonian (\ref{model_fulde})
that a gap $\Delta\sim (h_x)^\nu$ opens for small external magnetic
field, with an exponent $\nu$ depending
on the anisotropy. 
%
%
Our simulations for small anisotropy are not supporting
these predictions \cite{dmi02,essler_pri}. But due to the
difficulty to resolve accurately $\Delta_2$ using DMRG, we cannot
exclude completely an eventual small gap.
Although till now we are not able to give a clear answer to the gap-opening
question, but at least it seems that if there is a gap, the prefactor of the
scaling law would need to be very small and in any case too small to
explain experimental results on $\rm Yb_4As_3$.
\section{Staggered magnetic field}

\subsection{Model}
The staggered Dzyaloshinsky-Moriya interaction, which
is allowed \cite{aoki00,shiba00,oshikawa99}
in the $4f$-compound $\rm Yb_4As_3$, 
leads in an external homogeneous magnetic
field $h_z$ to an effective transversal
staggered field $h_x^{stag}$. In this context
the anisotropy is not relevant and we can
consider the (isotropic) Heisenberg case 
$J_{xy}=J_x\equiv J$:
\begin{equation}
H\,=\,J\sum_{i=1}^{L-1}\vec{S_i}\cdot\vec{S_{i+1}} 
+ h_x^{stag}\sum_{i=1}^L(-1)^{i}S_i^x -h_z\sum_{i=1}^LS_i^z~.
\label{H_stag}
\end{equation}
The above Hamiltonian is not invariant under reflection 
with respect to the mid point of the chain when $L$ is even. 
However in the standard implementation of the DMRG algorithm, 
$L$ is even and the reflection symmetry is used.
The Hamiltonian (\ref{H_stag}) can be easily made invariant 
under reflection by means of a local rotation, 
given by:
\[
\begin{array}{lcr}
 \left(
  \begin{array}{c}
   (-1)^i\,S_i^x\\
   (-1)^i\,S_i^y\\
   S^z_i\\
  \end{array}
 \right)
 & \rightarrow & 
 \left(
  \begin{array}{c}
   S_i^x\\
   S_i^y\\
   S^z_i\\
  \end{array}
 \right)\,.
\end{array}
\]
The transformed Hamiltonian reads:
\begin{eqnarray}
H &= & J\sum_{i=1}^{L-1}\big[\, S^z_i\,S^z_{i+1}  
-\left(S^x_i\,S^x_{i+1}\, +\, S^y_i\,S^y_{i+1}\right) 
                    \big] \nonumber\\
 &&\qquad +\, h_x^{stag}\sum_{i=1}^LS_i^x\, -\,h_z\sum_{i=1}^LS_i^z
\label{H_transformed}
\end{eqnarray}
%

\begin{figure}
\smallskip
  \begin{center}
  \includegraphics[width=7cm]{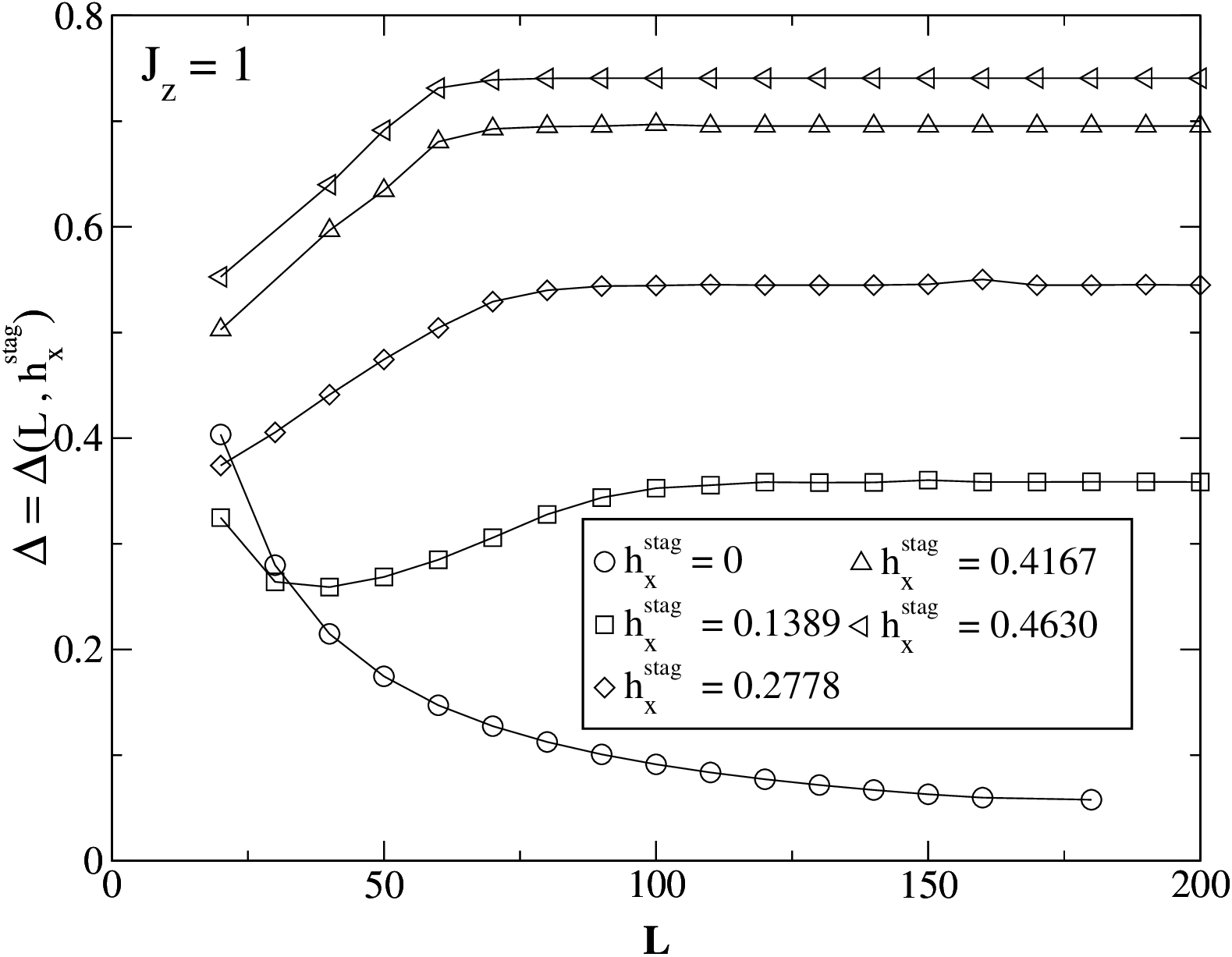}
  \end{center}
 \vspace{-0.5cm}
 \caption{DMRG results for the the gap for various staggered
 fields $h_x^{stag}$, as a function of chain length $L$,
 for $h_z=0$.
 } 
 \label{fig:gap_L}       
\end{figure}

\subsection{Effect of the staggered transversal field}
We start by analyzing the finite-size dependence
of the energy gap.
The Fig.~\ref{fig:gap_L} shows the behavior of the 
finite size system gap versus the staggered magnetic field.
One can clearly observe (i) that the gap vanishes 
in the thermodynamic limit only
for zero $h_x^{stag}$ and (ii) that the 
magnetic correlation length is large but finite
in the gapped case $h_x^{stag}>0$; the data for larger
system size $L$ is essentially flat for $h_x^{stag}>0$.

\begin{figure}
\smallskip
  \begin{center}
  \includegraphics[width=7cm]{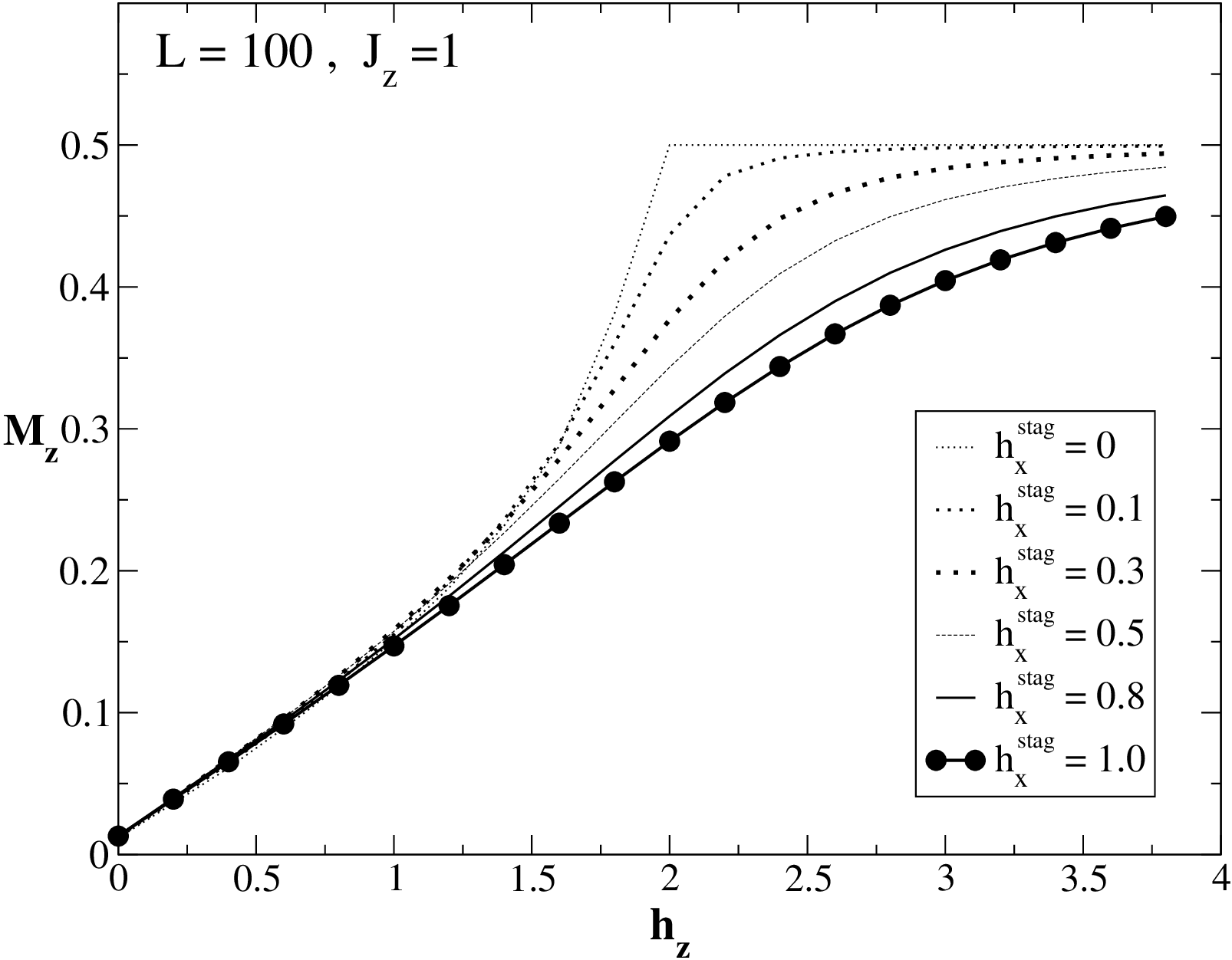}
  \end{center}
 \vspace{-0.5cm}
 \caption{Magnetization curves along the $z$-direction, 
          compare \cite{shibata01}.}
 \label{fig:Mz_stag}       
\end{figure}

In Fig.~\ref{fig:Mz_stag} we show the
magnetization along the $z$-direc\-tion as a function
of $h_z$, for various $h_x^{stag}$. Since
$J_{xy}=J_x\equiv1$ in (\ref{H_transformed}), these
results can be compared directly to those
given in Fig.~\ref{fig:Mx_L} for fields along $x$-direction.
We note that the second-order phase transition
to a completely magnetized state
occurring at $h_z=2$ in the absence of a staggered
field is progressively smeared out by $h_x^{stag}$;
the transversal field induces quantum fluctuations
into the magnetized state. 
These results confirm a similar study \cite{shibata01} 

\subsection{Comparison with experiment}

Now using the experimental estimation for exchange coupling 
in the isotropic Heisenberg chain 
$J\simeq 26\,K$, we fit the experimental data for the gap.
The staggered magnetic field 
$h_x^{stag}$ is proportional to the experimental magnetic field 
$H^{ext}$ via
$h_x^{stag}=c_0\,\sin(\theta)\,H^{ext}=c_0/g_\perp\sin(\theta)\,h_z$,
where $g_\perp$ is g-factor for a magnetic-field
perpendicular to the chain-direction. The g-factors
for $\rm Yb_4As_3$ are very anisotropic, $g_\perp$
has been estimated to be \cite{shibata01} 
$g_\perp\approx 1.3$.

\begin{figure}
\smallskip
  \begin{center}
  \includegraphics[width=8cm]{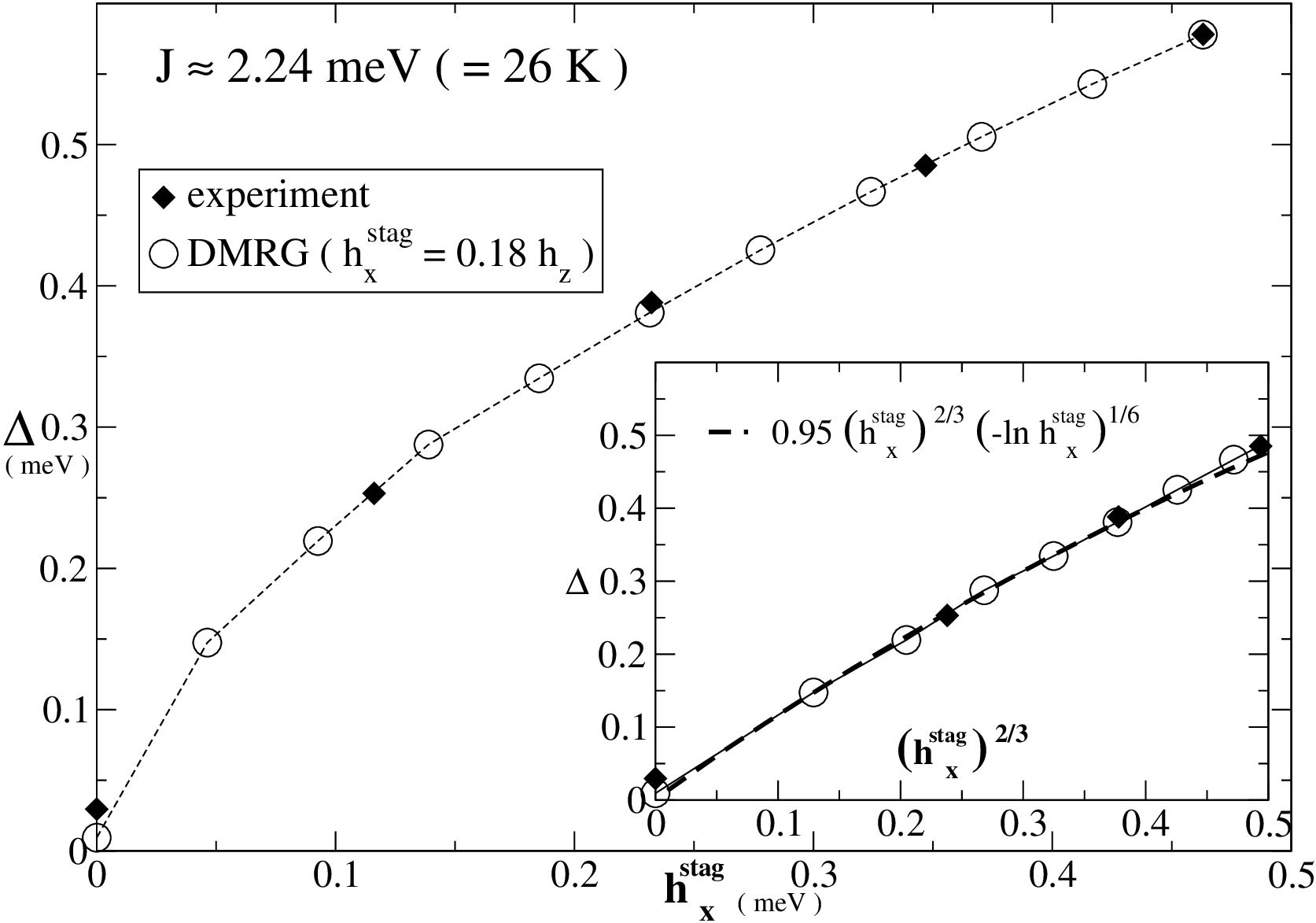}
  \end{center}
 \vspace{-0.5cm}
 \caption{Fit to experiment with data $(m = 70)$ extrapolated to the 
          thermodynamic limit.} 
 \label{fig:8}       
\end{figure}

Our fit yields $c_0/g_{\perp}\sin(\theta)\simeq0.18$.
The fitting to experimental data 
depends on $c_0$ which has not yet been determined precisely
by experiment.
While in the literature \cite{oshikawa99} 
$c_0\simeq0.27$, we get a good agreement with our DMRG 
simulations and the experimental curve for 
$ c_0\simeq0.23$ (assuming $g_\perp\simeq 1.3$)(Fig.\ref{fig:8}).

\subsection{General case} 
We consider now the case where the
staggered and the uniform magnetic field are not proportional.
We want to explore the excitation gap of
\begin{equation}
H =\sum_{i=1}^{L-1}(\vec{S_i}\cdot\vec{S_{i+1}}-\delta S_i^zS_{i+1}^z) 
     + h_x^{stag}\sum_{i}^L(-1)^{i}S_i^x -h_z\sum_{i}^{L}S_i^z~.
\label{H_general}
\end{equation}

\begin{figure}
\smallskip
  \begin{center}
  \includegraphics[width=8cm]{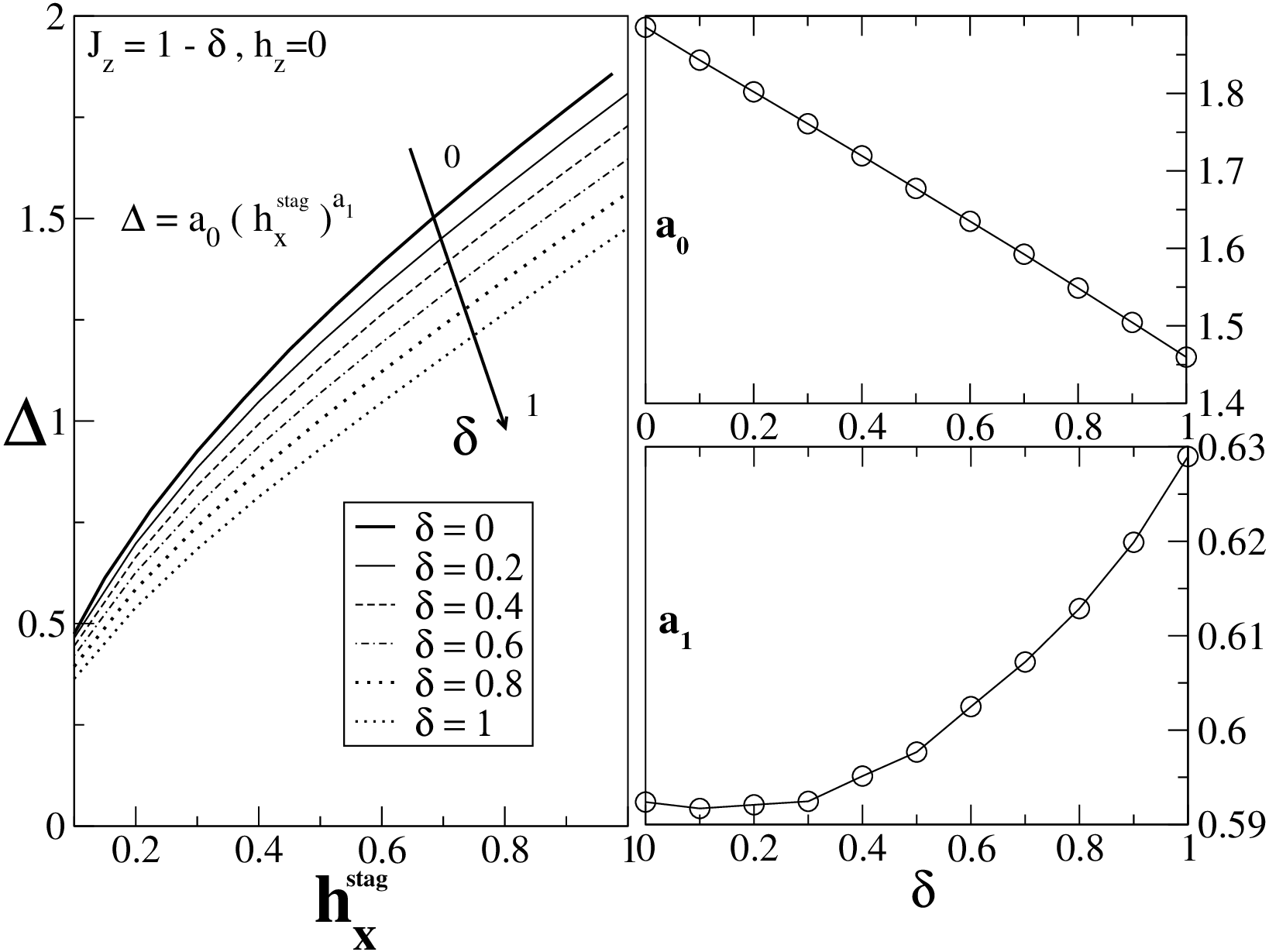}
  \end{center}
 \vspace{-0.5cm}
 \caption{Left panel: For $h_z=0$ the gap versus anisotropy $\delta$
 in the thermodynamic limit $(m = 40)$. Right panels: Dependence of the
 parameters $a_0$ and $a_1$ as a function of $\delta$. 
          } 
 \label{fig:fit_delta}       
\end{figure}

\begin{figure}
\smallskip
  \begin{center}
  \includegraphics[width=8cm]{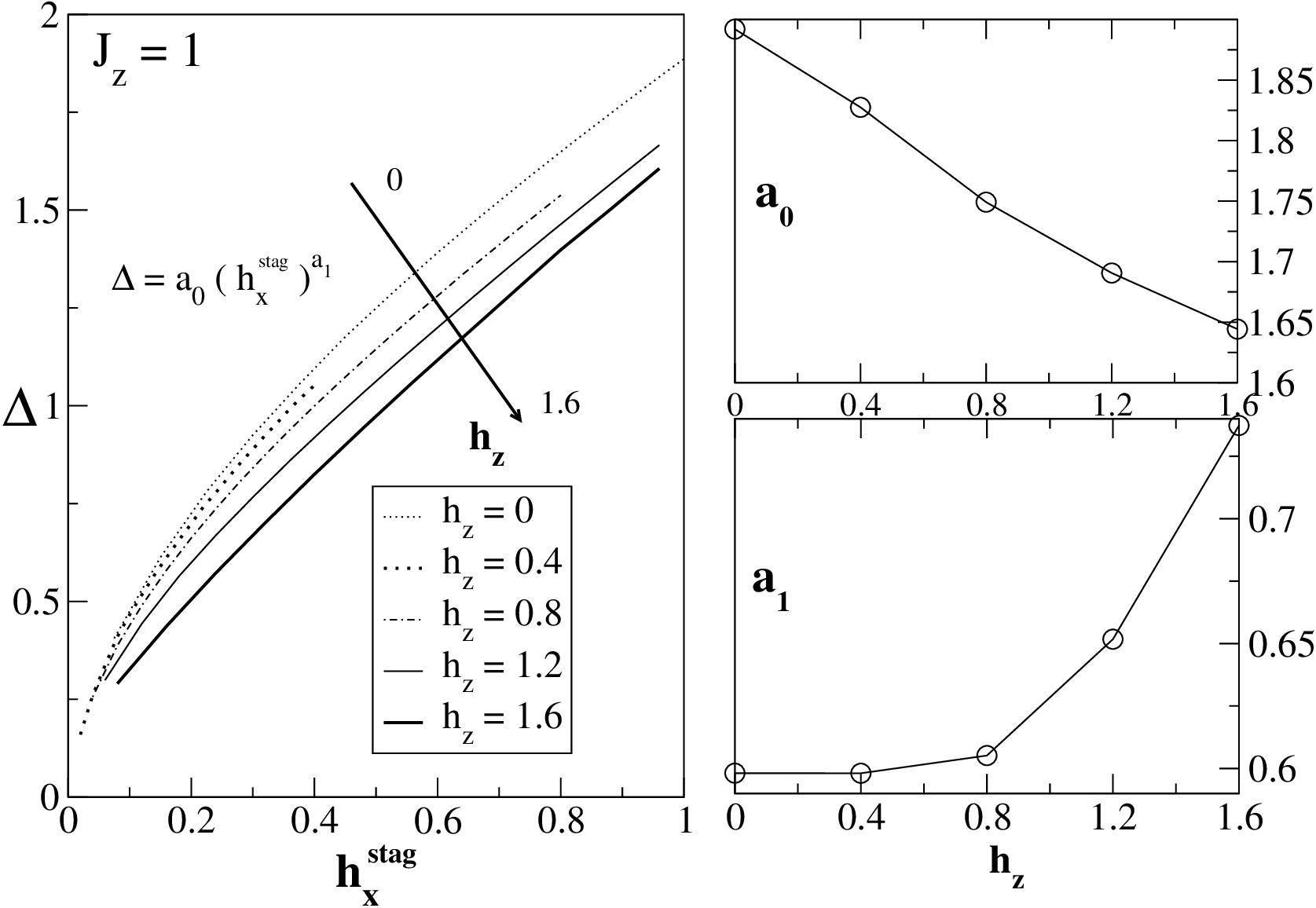}
  \end{center}
 \vspace{-0.5cm}
 \caption{Left panel: For $\delta=0$ the gap versus uniform
          longitudinal field $h_z$
 in the thermodynamic limit $(m = 40)$. Right panels: Dependence of the
 parameters $a_0$ and $a_1$ as a function of $\delta$. 
          } 
 \label{fig:fit_stag}       
\end{figure}

In Fig.~\ref{fig:fit_delta} we present for $h_z=0$
the gap $\Delta$ as
a function of the anisotropy $\delta$ and the
parameters $a_0$ and $a_1$ entering in the scaling-law
\begin{equation}
\Delta \ =\ a_0\left( h_x^{stag}\right)^{a_1}
\label{fit}
\end{equation}
for the gap. Bosonization predicts \cite{oshikawa97,affleck99} 
$a_0\approx1.85$ and $a_1=2/3$ for the
isotropic case ($\delta=0$, $h_z=0$) for 
small staggered magnetic field $h_x^{stag}$.
For the isotropic case $(\delta=0$) 
we present in Fig.~\ref{fig:fit_stag} the same data as
a function of $h_z$.

We also examined the DMRG-data using (\ref{fit})
including multiplicative logarithmic corrections
\cite{oshikawa97,affleck99}:
$\Delta \ =\ a_0\left( h_x^{stag}\right)^{a_1}\left|\log h_x^{stag}\right|^{1/6}$
We found essentially the same values for the parameters
$a_0$ and $a_1$ as presented in Fig.~\ref{fig:fit_delta}
and Fig.~\ref{fig:fit_stag}.
We note, that (\ref{fit}) hold only in the asymptotic limit
$h_x^{stag}\to0$, which we do not examine in the present
study. The results presented in  Fig.~\ref{fig:fit_delta}
and Fig.~\ref{fig:fit_stag} show $a_0$ and $a_1$ as obtained
for overall fits to the gap, for $h_x^{stag}\le0.8$. We 
believe this parameter-region to be experimentally relevant.

\section{Conclusions}

We have studied the anisotropic Heisenberg-chain
with staggered and uniform transversal and uniform longitudinal
field by DMRG. We found no evidence 
for a substantial gap opening for a homogeneous transversal field,
as predicted by a mean-field \cite{uimin00} and
a scaling \cite{dmi02} analysis. We found, however,
a gap opening for a staggered transversal field,
in accordance with previous studies \cite{oshikawa97,affleck99}.
These results lead to the conclusion, that
the gap-opening in the
the rare-earth, $4f$-compound $\rm Yb_4As_3$
compound in an external magnetic field is
attributable to a staggered Dzyaloshinsky-Moriya 
tensor \cite{oshikawa99,aoki00,shiba00}.


\end{document}